\documentclass[%
aps,
twocolumn,
superscriptaddress,
nofootinbib,
 amsmath,amssymb,
]{revtex4}

\usepackage{graphicx}
\usepackage{dcolumn}
\usepackage{bm}
\usepackage{xcolor}
\usepackage{upgreek}
\usepackage{soul}
\usepackage{hyperref}
\usepackage{comment}
\hypersetup{
	colorlinks = true,
	pdftitle = {},
	pdfauthor = {},
	pdfkeywords = {},
	linkcolor = blue,
	citecolor = blue,
	filecolor = black,
	urlcolor = blue
}
\begin{document}

\title{A New Resolution Odyssey in Electron Microscopy}
\author{JuanCarlos Idrobo}
\email[Corresponding author: ]{idrobojc@ornl.gov}

\affiliation{Center for Nanophase Materials Sciences, Oak Ridge National Laboratory, Oak Ridge, Tennessee 37831, USA}

\date{\today}
\maketitle

In 1959, Richard Feynman in his now widely-known lecture to the American Physical Society, “Plenty of Room at the Bottom,” argued that many fundamental biological questions could be simply answered by looking “at the thing” if one could make the electron microscope 100 times better than what was available at the time \cite{r1}.  He challenged his audience to improve the resolution of the electron microscope from roughly 1 nanometer [nm] to 0.01 nm (or 10 picometer [pm]).  He argued, correctly, as theoretically shown by Otto Scherzer almost 10 years earlier \cite{r2}, that the path to improve the resolution of the electron microscope was to use electron lenses that did not necessarily maintain rotational (cylindrical) symmetry.

It took the scientific community almost another 40 years to show experimentally that precisely combining a suitable sequence of non-cylindrically symmetric electron lenses can result in an improvement of the resolution of the electron microscope \cite{3,4}.  This achievement, known in the microscopy community as aberration correction, has simply revolutionized how we study materials and biological matter at the atomic level \cite{5}. 

Yet, the challenge presented by Feynman of achieving a 10 pm spatial resolution has not been reached. We are close.  The constant improvement of aberration correction optics, driven by more stable electronics, has recently (as of 2020) resulted in images with spatial resolutions of about 40 pm \cite{6}.  If this trend continues, there is little doubt that microscopes will eventually have the imaging power envisioned by Feynman.

Besides aberration correction, there have been many other improvements in electron microscopy, which collectively allow us to study matter in ways that are not simply given by having a microscope with better spatial resolution.

One clear example is the recent reemergence of monochromation \cite{7}.  The driving force behind monochromation is the need to obtain electron energy spectroscopy at higher energy resolution.  The key here, however, is to achieve the enhancement of the energy resolution while still maintaining a high spatial resolution.  Monochromation in this context has been available since the early 2000s \cite{8}. However, only since 2014 has the energy resolution reached the point where infrared optical excitations and phonons are resolved in the electron microscope, even reaching atomic resolution \cite{9,10}.  This development has opened a new door to study optical excitations, thermal and phonon properties of materials at a new fundamental level \cite{11,12,13,14}.

Another recent example is driven by the need to image magnetic materials at very-high resolution in field-free imaging modes \cite{15}. This development required a new design of electron lenses in the pole piece of the microscope, which is where the samples are located.  Due to this development, it is now possible to resolve individual atomic columns of materials without the necessity of the usual magnetic field ($\sim 1-2$ Tesla) present on the sample to keep the electrons correctly focused.

There have been other developments that are equally important, such as better digital detectors (cameras), and stages to perform a large variety of in situ and operando experiments, as well as the capability to image materials at cryo temperatures (using liquid nitrogen as a coolant), albeit with a moderately reduced stage stability.

However, the key question that microscopists, materials scientists, physicists, chemists and engineers are asking now is: What is the next big frontier in electron microscopy, and what are the fundamental scientific problems, technological leaps and benefits to society that can be addressed after crossing that frontier?  We can easily ask this question to ten different people and get eleven different answers.  So, instead of asking the question directly, I will try to follow Richard Feynman's lead and instead present a simple statement of a dream and see where it leads us, what is required to achieve it, and how many new doors this dream might open up.

Well, I dream of the time where the magnetic moments of individual atoms can be resolved in an electron microscope as easily as we can now resolve and distinguish individual atoms of different elements. Admittedly, this feat is not always easy, but can regularly be achieved in solid samples.

\begin{figure*}[ht]
\includegraphics[width=0.78\textwidth]{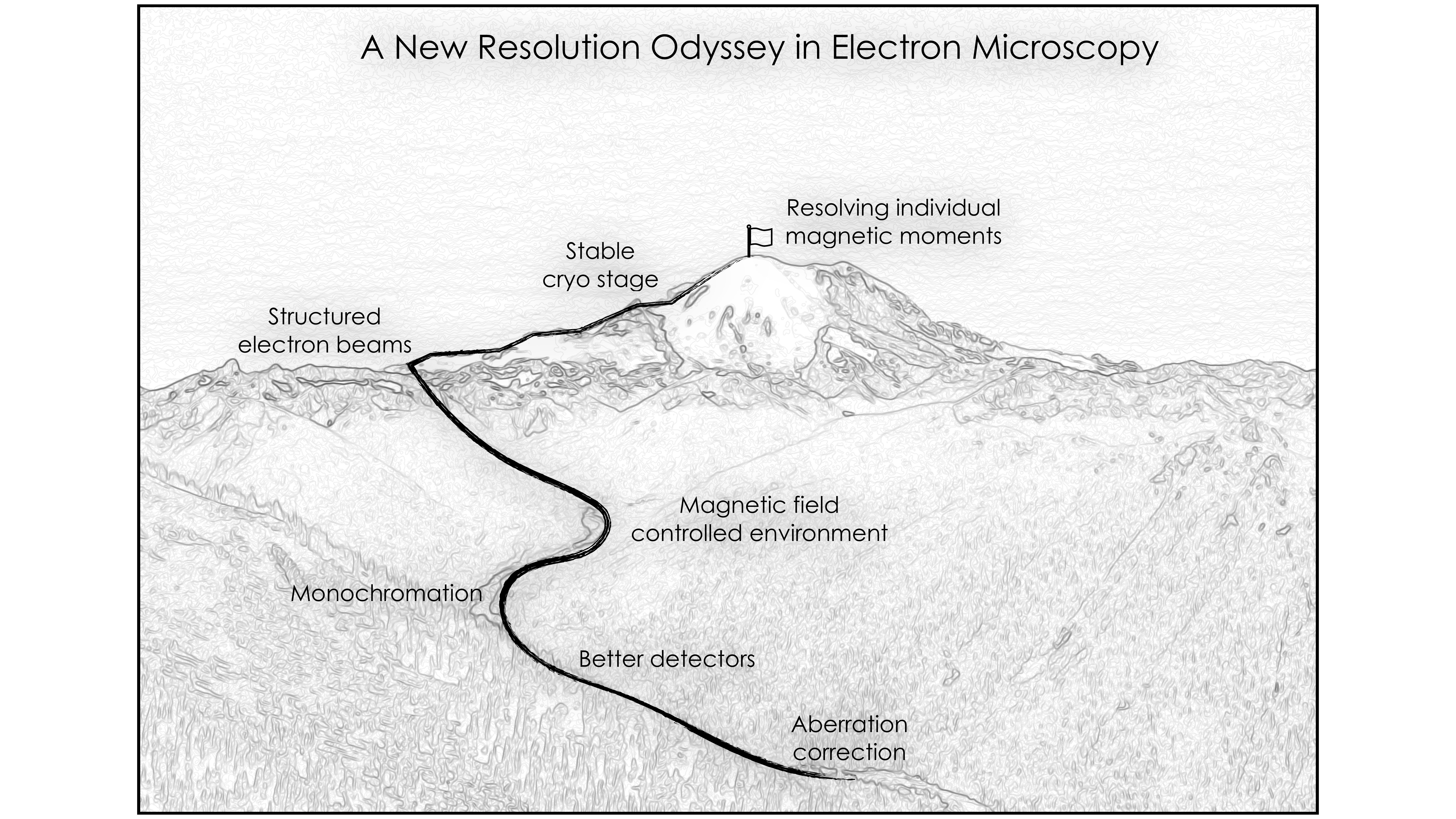}
\end{figure*}

What does it mean to resolve the magnetic moments of individual atoms? It means, for instance, that if I see two nearby iron atoms in a material, I should be able to quantify via electron spectroscopy the nature of their magnetic moments \cite{16}; how much is coming from spin and how much is arising from electron orbitals.  I could also answer if the two iron magnetic moments are aligned parallel (ferromagnetically), antiparallel (anti-ferromagnetically), or in some other order.

For the previous experiment to become a reality we need to be able to control the phase of the incoming electron wave better than is now possible.  We would require the incoming electrons to have a helical or twisted wavefront (known in optics as vortex beams), which imprints an orbital angular momentum proportional to the winding of the helicity \cite{17,18}.  And of course, these vortex beams need to be of atomic size if we want to resolve individual atoms or atomic columns \cite{19}.

Expanding the previous idea to many more atoms means that we should be able to directly image in materials all kind of magnetic phases, domains, boundaries, etc.

If we could reduce the temperature of the sample far below liquid nitrogen, let us say to about 1 K, or even to 10 K, then a complete new panorama opens up. At those low new temperatures we could map how magnetism emerges at the atomic scale and how phonons (at different temperatures) control the extent of those magnetic correlations.  For that to be a reality, we would also need to use the new pole piece mentioned before, where the external magnetic field at the sample volume can be nulled or varied in a controlled manner without reducing the spatial resolution of the microscope. 

We could also study the emergence of superconductivity at the atomic scale and how magnetism hinders or enhances different superconducting materials.  

There are also more exciting experiments that require a bit more of our imagination.  For example, electron vortex beams carrying orbital angular momentum could be used as a pump probe, similarly to circularly polarized light \cite{20}, to activate anomalous Hall currents in topological materials that could be measured -- with contacts made previously on the sample via current nanofabrication techniques.  However, the advantage here is that these electron vortex beams are of atomic size, such that atomic structural information of the samples can also be obtained simultaneously.  In other words, we could map and activate at the atomic scale the Hall effects of materials, and see how defects or interfaces affect their topological properties.

Another advantage of using electron vortex beams is that we are not restricted to only two angular momentum values as occurs with circular or linear polarized light.  We could imprint larger values of orbital angular momentum, albeit with a degradation of spatial resolution, and see effects that are beyond single photon emissions.

The electron microscope could also be used in a diffraction mode, where momentum dispersions can be resolved.  In this mode, we could use monochromated electron (vortex  and conventional) beams to measure the dispersions of magnons, excitons, phonons, and polaritons in materials such as superconductors, topological insulators, and van der Waals hetero-structures to name only a few. 

Monochromated electron vortex beams should enable us to study valley states, and with enough energy resolution (currently as 2020 is 3 meV at a 20 kV acceleration voltage \cite{21}) to unveil spin orbit coupling, moire potentials as well as other electron correlations at similar energy ranges.  Further improvements in energy resolution should perhaps allow us to see superconducting gaps emerging at critical temperatures, and directly correlate them with the phonon properties of the superconductor. We might be able to achieve a whole new level of understanding of superconductivity.

We should not only be limited to conventional electron energy-loss spectroscopy experiments.  We could also spectroscopically resolve the orbital angular momentum absorbed by the materials in real and momentum space \cite{22}, giving unique information about the electronic correlations dominating the properties of materials. 

Here, I am just touching the surface of many new experiments that will be possible if we venture to develop the technology required to resolve the magnetic moments of individual atoms.  Further experiments and the information that we will obtain from them are limited only by our imagination. 

But you might wonder now, what do we need to accomplish such dream? Well, we need development on many fronts simultaneously.  

First, aberration correction and monochromation need to be further developed, where the emphasis is not only on spatial and energy resolution improvements, but also on long-term stability, ease of use and with a range of automated procedures.  Software that allows fully remote operation of these new electron microscopes, where monochromation and aberration correction is the base line, should be thought out from the ground up.

We definitely need a revolution in stage stability at cryogenic temperatures. The majority of the sample stages currently available, even in state-of-the-art electron microscopes, were designed over 30 years ago without much improvement since.  We cannot expect a different outcome in stage stability if we just repackage designs from the last century. 

If the Laser Interferometer Gravitational-Wave Observatory (LIGO) is capable of detecting a change of less than one ten-thousandth the charge diameter of a proton \cite{23}, with the stringent stability conditions that the feat implies, why cannot we have a low-temperature (1-10 K) cryo stage that effectively drifts by less than 10 pm in 10 minutes or so? 

LIGO can mitigate vibrations of big masses (of hundreds of kilograms) to the order of $10^{-19}$ meters to detect gravitational waves \cite{23}. Perhaps we can have a cryo stage in our new electron microscopes that can achieve just a 100-millionth of what LIGO does.  Obviously, I am not asking for an overnight solution. But we need to walk away from old stage designs and move towards stability, stability and stability!

As mentioned previously, we also need to imprint orbital angular momentum of the incoming electrons.  Or in more general terms, we should be able to produce structured electron beams, where phase and shape can be carefully tuned.  We also need further development of electron spectroscopy of orbital angular momentum. 

We need to control the environment of the sample, for instance, its external magnetic field, as mentioned previously.  Biasing as well as heating the sample, combined with the capability for light illumination (and measuring the outgoing light) will be required.  We could also have a different line of stages, where the emphasis is on other external variables such as the control of gases and liquids, addressing then the requirements for novel catalyst and battery experiments, to just mention two straightforward examples.

We have seen the revolution that the new direct electron detectors caused in structural biology \cite{24}.  We need cameras capable of detecting single electrons while maintaining a high-dynamic range without saturation in the new microscopes. They are critical for all the spectroscopy experiments mentioned previously, as well as for differential phase contrast and ptychographic imaging methods used to reveal local electric fields, magnetic fields, charge density of materials and even to correct a posteriori aberrations of the incoming electron wave. 

A point that I have not yet touched directly is the incorporation of novel data analysis algorithms, such as Artificial Intelligence and Machine Learning. 

I sometimes like to use a mountain climbing analogy when I talk about these topics to a general audience. One can compare aberration-correction to getting to a base camp on a mountain, monochromation to a second camp, a variable magnetic field pole piece to a third camp, and so on for a new lens or controllable angular momentum, with various essential microscope parts perhaps analogous to rock-climbing equipment \cite{25}.

Based on that analogy, if we would then like to climb Mount Everest, after years of training and conditioning, perhaps we could do it without an oxygen mask, but the risk is that we might never get to the top of the mountain or might be even more likely die in the attempt.  Modern data analytics here plays the same role as the oxygen mask in my example.  We need to incorporate data analytics throughout the whole experimental process.

Another very important variable that I have also not touched on so far is time.  Imagine that besides resolving the magnetic moments of individual atoms, we could also see their dynamics! In a nutshell, the conditions required to achieve such ultra-fast experiments in the electron microscope are obtained by producing single-electron packets emitting from an electron cathode using femtosecond optical pulses.

Using these novel electron emitters, we should be able to study spin relaxation times, and study the coherence and the entanglement of quantum topological states, to list three very simple examples.  The experiments could be performed stroboscopically to increase signal-to-noise ratios in the images and spectra, assuming the phenomena to be studied can be cyclically controlled.  We could also reduce the speed of the optical pulses and play with external variables such as temperature or biasing to study the dynamics of ionic movement in batteries and fuel cells.

Finally, we need to address the technological leaps and social benefits that would result from this adventure to resolve the magnetic moments of individual atoms in the electron microscope.  Here, I think any example would actually fall short, so I am going to briefly illustrate a single case.  

If we could figure out a way to create strong magnets that do not rely on rare-earth elements, we could decrease significantly the cost of many electronic devices, ranging from the small haptic engines used in smart watches, to household air conditioning units, to electric motors in electric cars and wind turbines.  This will also result in a quantifiable environmental benefit by reducing the mining of rare-earth elements and their chemical separation. 

But, from my perspective, the most exciting thing is not simply the developments in materials and the improvement of their properties for our current technology.  I am way more excited about the unknown unknowns, for those discoveries that are awaiting us, for that future technology that will emerge when we discover new physical phenomena and better understand how matter behaves.   

\begin{acknowledgments}
This research was conducted at the Center for Nanophase Materials Sciences, which is a DOE Office of Science User Facility.  The author would like to thank to Ondrej Krivanek, Niklas Dellby, J\'{a}n Rusz, Stephen Pennycook, Andrea Kone\v{c}n\'{a}, Robert Klie, Rafal Dunin-Borkowski, Naoya Shibata, Ryo Ishikawa and Andrew Lupini for the countless stimulating discussions and emails through many years about some of the topics covered in this essay. 
\end{acknowledgments}
{\color{gray}
\section*{Copyright notice}
 This manuscript has been authored by UT-Battelle, LLC under Contract No. DE- AC05-00OR22725 with the U.S. Department of Energy. The United States Government retains and the publisher, by accepting the article for publication, acknowledges that the United States Government retains a non-exclusive, paid-up, irrevocable, world-wide license to publish or reproduce the published form of this manuscript, or allow others to do so, for United States Government purposes. The Department of Energy will provide public access to these results of federally sponsored research in accordance with the DOE Public Access Plan (http://energy.gov/downloads/doe-public-access-plan).}


\begin{thebibliography}{99}
\bibitem{r1} R.P. Feynman,\textit{ Engineering and Science} 22–36 (February 1960).
\bibitem{r2} O. Scherzer, \textit{Optik} \textbf{2}, 114 (1947).
\bibitem{3}	J. Zach, and M. Haider, \textit{Optik} \textbf{93}, 112-118 (1995).
\bibitem{4}	O.L. Krivanek, et al.,\textit{ IoP Conference Series} \textbf{153},  35 (1997).
\bibitem{5}	H. Rose, M. Haider, K. Urban, and O.L. Krivanek, recipients of the 2020 Kavli Prize in Nanoscience ``\textit{for sub-ångström resolution imaging and chemical analysis using electron beams}," (2020).
\bibitem{6}	S. Morishita et al., \textit{Microscopy} \textbf{67}, 46 (2018).
\bibitem{7}	O.L. Krivanek, et al., \textit{Nature} \textbf{514}, 209–212 (2014).
\bibitem{8}	P.C. Tiemeijer, \textit{Proc. EMAG ’99, Sheffield. Institute of Physics Conference Series} \textbf{161}, 191–194 (1999).
\bibitem{9}	K. Venkatraman, et al., \textit{Nat. Phys.} \textbf{15}, 1237–1241 (2019).
\bibitem{10}	F. Hage, et al., \textit{Science} \textbf{367}, 1124-1127 (2020).
\bibitem{11}	M.L. Lagos, et al., \textit{Nature} \textbf{543}, 529–532 (2017).
\bibitem{12}	J.C. Idrobo, et al., \textit{Phys. Rev. Lett.} \textbf{120}, 09590 (2018).
\bibitem{13}	J.A. Hachtel, et al., \textit{Science} \textbf{363}, 525–528 (2019). 
\bibitem{14}	R. Senga et al., \textit{Nature} \textbf{573}, 247–250 (2019).
\bibitem{15}	N. Shibata, et al., \textit{Nat. Comm.} \textbf{10}, 2308 (2019).
\bibitem{16}	C. Hébert and P. Schattschneider, \textit{Ultramicroscopy} \textbf{96}, 463-468 (2003).
\bibitem{17}	M. Uchida and A. Tonomura, \textit{Nature} \textbf{464}, 737–739, (2010).
\bibitem{18}	J. Verbeeck, H. Tian, and P. Schattschneider, \textit{Nature} \textbf{467}, 301–304, (2010).
\bibitem{19} J. Rusz and S. Bhowmick, \textit{Phys. Rev. Lett.} \textbf{111}, 105504 (2013).
\bibitem{20}	F.K. Mak et al., \textit{Science} \textbf{344}, 1489-1492 (2014).
\bibitem{21}	N. Dellby et al., \textit{Micros. and Microanal.} 2020 \textit{Proceedings}, 703 (2020).
\bibitem{22} V. Grillo et al.,\textit{ Nat. Comm.} \textbf{8}, 15536 (2017).
\bibitem{23} B.P. Abbott et al., \textit{Phys. Rev. Lett.} \textbf{116}, 061102 (2016).
\bibitem{24} A. R. Faruqi, R. Henderson, \textit{Curr. Opin. Struct. Biol.} \textbf{17}, 549 (2007).
\bibitem{25} Image of Mount Rainer in WA USA, originally taken by the author in August $14^{th}$ 2016.

\end{thebibliography}
\end{document}